\documentclass[twocolumn,showpacs,preprintnumbers,amsmath,amssymb,floatfix]{revtex4}
\usepackage{graphicx}
\usepackage{dcolumn}
\usepackage{bm}
\begin{document}

\preprint{PRD}

\title[LISA optical links]{Relativistic analysis of the LISA long range optical links}

\author{Bertrand Chauvineau, Tania Regimbau and Jean-Yves Vinet}

\affiliation{Department ARTEMIS \\ Observatoire de la C\^ote d'Azur, \\ BP 429
06304 Nice (France). }

\author{Sophie Pireaux}
\affiliation{Department GEMINI \\Observatoire de la C\^ote d'Azur, \\ Avenue Copernic
06130 Grasse (France). }

\date{\today}

\begin{abstract}
The joint ESA/NASA LISA mission consists in three spacecraft on heliocentric orbits, 
flying in a triangular formation of 5 Mkm each side, linked by infrared optical beams. 
The aim of the mission is to detect gravitational waves in a low frequency band. 
For properly processing the scientific data, the propagation delays between spacecraft
must be accurately known. 
We thus analyse the propagation of light between spacecraft in order to systematically derive 
the relativistic effects due to the static curvature of the Schwarzschild 
space-time in which the spacecraft are orbiting with time-varying light-distances.
In particular, our analysis allows to evaluate rigorously the Sagnac effect, 
and the gravitational (Einstein) redshift.
\end{abstract}


\maketitle

\section{\label{intro}Introduction}
The three spacecraft forming the LISA constellation have a complex motion (see \cite{DNV} 
for instance) allowing the triangular constellation to remain approximately rigid during 
the annual revolution. There is simultaneously a rotation of the triangle around its center, 
and the orbital motion of its center. A number of papers have considered the effect of the 
rotation on light propagation. Accurate knowledge of light propagation delays are very 
important for implementing the TDI (Time-Delay Interferometry) technique (see \cite{TEA},
\cite{NPDV}) which is mandatory for eliminating laser phase and similar noises. In particular, in 
a simulation code, it is necessary to generate time delays as realistic as possible. 
This is why we have carried out a full general relativistic treatment of light propagation 
in the gravitational potential of the Sun. The background metric created by the 
Sun is responsible for both the orbits, and the fine structure of time delays. 
Even if the orbit of any spacecraft is taken as purely Keplerian, the time interval 
between the emission of a photon off spacecraft A and its detection in spacecraft B 
must be thoroughly investigated. In particular, the motion of the target B during the 
interception must be evaluated with an accuracy level consistent with that of the null 
geodesic followed by the photon.
We therefore propose an expansion of both the null geodesic equation from A to B and 
the motion of the target B, in powers of the small dimensionless parameter
\begin{equation}
\epsilon=r_S/2r=GM/(rc^2)
\label{epsilondef}
\end{equation}
where $r_S=2GM/c^2$ is the Schwarzschild radius of the Sun (of mass $M$) and $r$ 
the coordinate distance of the spacecraft from the Sun. We intend to derive the time 
delay between spacecraft and the global frequency shift (including classical Doppler
due to relative motion or orbital shift plus Einstein shift). The first quantity is 
essential for TDI, the second seems important for technology, because large 
frequency shifts, variable in time over a year induce requirements on the
ultrastable oscillators used for compensation of orbital frequency shifts.
We propose formulas for computing the time delays necessary for a realistic
numerical model and we show that in the global frequency shift, the overall 
general relativistic contribution (first order terms) 
is much smaller than a naive 
estimate would suggest.

The method we present is based on a matching of geodesic equations of spacecrafts A and B, 
with the (null) geodesic equation of the photon emitted by A and received by B. 
\section{Computation of (coordinate-)time transfer in the literature}
Different methods have been considered for computing the flight time
of photons and the
gravitational frequency shift between two space-time points $x_{A}^{\alpha
}\equiv \left( t_{A},\mathbf{x}_{A}(t_{A})\right) $ (emitter) and $%
x_{B}^{\alpha }\equiv \left( t_{B},\mathbf{x}_{B}(t_{B})\right) $
(receiver). 
An alternative to integrating the geodesic equations of motion
is to use the world function $\Omega \left( x_{A}^{\alpha },x_{B}^{\alpha
}\right) $ developed by Ruse \cite{R} and Synge \cite{S}, defined as the
squared geodesic distance between two space-time points with $\Omega \left(
x_{A}^{\alpha },x^{\alpha }\right) =0$ being the equation of the light cone
at $x_{A}^{\alpha }$. The time delay corresponds to the future cone
equation, and is given by $\tau \left( t_{A},\mathbf{x}_{A}(t_{A}),\mathbf{x}%
_{B}(t_{B})\right) \equiv t_{B}-t_{A}=R_{AB}/c-\stackrel{(PN)}{\Omega }%
\left( t_{A},\mathbf{x}_{A}(t_{A}),\mathbf{x}_{B}(t_{B})\right) /(c$ $R_{AB})
$, the subscript PN meaning Post-Newtonian contributions, $\mathbf{R}%
_{AB}\equiv \mathbf{x}_{B}(t_{B})-\mathbf{x}_{A}(t_{A})$ and ${R}_{AB}\equiv
\left| \mathbf{R}_{AB}\right| $.

Linet et al (2002) \cite{LT} provide the world function, $\Omega
\left( t_{A},\mathbf{x}_{A}(t_{A}),\mathbf{x}_{B}(t_{B})\right) $, up to
order $1/c^{3}$ in the full Nordtvedt-Will Parametrized PN (PPN) formalism
(10 parameters characterizing alternative theories of gravitation, including
the usual PN parameters $\beta $ and $\gamma $). Their results yield an
expression for the coordinate time transfer, $\tau \left( t_{A},\mathbf{x}%
_{A}(t_{A}),\mathbf{x}_{B}(t_{\mathbf{B}})\right) $, at order $1/c^{4}$. The
authors also provide $\tau \left( t_{A},\mathbf{x}_{A}(t_{A}),\mathbf{x}%
_{B}(t_{\mathbf{A}})\right) $ which contains explicitly the motion of the
receiver and the corresponding Sagnac effects of orders $1/c^{2}$, $1/c^{3}$%
, $1/c^{4}$.
They applied the earlier expression to an isolated, axisymmetric rotating
body, assuming a stationary gravitational field and a constant velocity of
the central gravitational body with respect to the universe rest frame.
Their systematic procedure to compute multipole moment contributions in $%
\Omega \left( \mathbf{x}_{A}(t_{A}),\mathbf{x}_{B}(t_{B})\right) $ and $\tau
\left( \mathbf{x}_{A}(t_{A}),\mathbf{x}_{B}(t_{B})\right) $ was then
particularized to obtain the explicit contributions of the mass monopole,
mass quadrupole moment and of the intrinsic angular momentum of the rotating
gravitational body. Restricting to fully conservative metric theories of
gravitation, without preferred location effects (that is all PPN parameters
vanish except $\beta $ and $\gamma $), they determined the fractional
frequency shift up to order $1/c^{4}$.
The authors then performed numerical estimates of the frequency shifts in
the gravitational field of the Earth. They assumed $A$ on board the
International Space Station (ISS), orbiting at an altitude of 400km, and $B$
in a terrestrial station; as this is the case of ESA's Atomic Clock Ensemble
in Space (ACES) mission, planned for 2009-2010, aiming at an accuracy of $%
10^{-16}$ in fractional frequency. The formula the authors give yields all
the gravitational corrections to frequency shifts up to $10^{-18}$ in the
vicinity of the Earth.

The above cited paper in fact generalizes the work of Blanchet et al
(2001) \cite{BSTW}, based on the geodesic approach, who provided the time
transfer and frequency shifts up to order $1/c^{3}$ in the setting of
General Relativity (where $\beta =\gamma =1$ are the only nonvanishing PPN
parameters), for a monopole mass, without intrinsic spin. Paper \cite{BSTW}
considered the two-way laser link in addition to the one-way laser link
presented in paper \cite{LT}. However, for the LISA mission, only single
links are pertinent since laser light is not reflected from one station to
another, LISA\ consisting in three pairs of laser links.

Le Poncin-Lafitte et al (2004) \cite{PLT} present a generic procedure based on
the world function to provide, in an iterative way, the propagation time of
a photon between space-time events $x_{A}^{\alpha }$ and $x_{B}^{\alpha }\ $%
at the $n$th post-Minkowskian order (developpement in powers of $G$) from
the contributions of orders $p=n-1$, $n-2,...,$ $1$, $0$, i.e. of $\stackrel{%
(p)}{\Omega }\left( t_{A},\mathbf{x}_{A}(t_{A}),\mathbf{x}_{B}(t_{B})\right) 
$. The authors then apply their formalism to a static spherically symmetric
space-time within the second post-Minkowskian approximation, with the usual
post-Newtonian parameters $\beta $, $\gamma $ plus parameter $\delta $ for
the $G^{2}$-term in $g_{ij}$ (normalized so that in General Relativity $%
\delta =1$). Within those assumptions, they obtain the expressions of the
world function and time transfer in the case of a simple monopolar and
nonrotating body, up to order $G^{2}$ (containing order $1/c^{5}$). The
motion of the stations (and corresponding Sagnac effects) is not explicitly
considered there. 

We are interested in the time variations of the 
interspacecraft propagation delays over a year for spacecraft having 
very special orbits, this is the reason why we develop a special approach,
on principles analogous to the preceding works, but in a form more suitable
to our purpose. 
\section{Post-Newtonian Theory}
\subsection{Notations}
Coordinates $(ct,x,y,z)$ are denoted by $x^\alpha$. We use the notation $r=\sqrt{x^ix^i}$, 
where latin indices take only the spatial values 1, 2 and 3. Then, $r$ cannot be understood 
as a physical distance in all the steps of the developments. It should be referred to as a 
coordinate distance.

Four-velocities are written $u^\alpha\equiv dx^\alpha/d\tau$, where $\tau$ is the proper time, 
given by $c^2d\tau^2=-ds^2$, since the signature chosen for the metric is $(-,+,+,+)$. 
We frequently use the three-velocity $v^i\equiv dx^i/dt$. This quantity cannot be understood 
as a physical velocity in all the steps of the developments.
When a quantity $Q$ is evaluated up to order $p$ (integer or half-integer) in $\epsilon$, 
it will be written 
$$\stackrel{(\rightarrow p)}{Q}=\stackrel{(0)}{Q}+\stackrel{(1/2)}{Q}+...+\stackrel{(p)}{Q}$$
where $\stackrel{(l)}{Q}\ $ is the contribution of order $l$.

The notation ${\bf U}$ represents the three numbers $U^i$, and ${\bf U.V}$ represents the 
sum $U^iV^i\equiv U^1U^1+U^2U^2+U^3U^3$. The notation ${\bf U}^2={\bf U.U}$ will also be used.
$\eta_{\alpha\beta}=(-1,+1,+1,+1)$ is the Minkowski metric.
 
In this paper, we derive the (relative) frequency shift of a photon, linking 
an emitter spacecraft A to a receiver spacecraft B, up to order $3/2$ in $\epsilon$.
\subsection{Generalities on the metric}
We consider a metric of the generic static form
\begin{equation}
ds^2=-I(\epsilon)c^2dt^2+J(\epsilon)\delta_{ij}dx^idx^j
\label{metric}
\end{equation}
where $I$ and $J$ are two functions depending on the theory (e.g. General Relativity, 
Scalar-Tensor theory,...). Both are assumed expansions in integer powers of $\epsilon$. 
The leading order in $v^i/c$, is 1/2 (Kepler). A photon worldline, described by the wave 
vector $k^\alpha\equiv dx^\alpha/d\lambda$, where $\lambda$ is an (arbitrary) affine 
parameter, obeys the isotropic condition 
\begin{equation}
k_\alpha k^\alpha=0 
\label{isotrop}
\end{equation}
and the null geodesic equation
\begin{equation}
\frac{dk^\alpha}{d\lambda}+\Gamma^\alpha_{\mu \nu}\,k^\mu\,k^\nu \ = \ 0
\label{geophot1}
\end{equation}
or, as well
\begin{equation}
\frac{dk_\alpha}{d\lambda}= \frac{1}{2}\,\,k^\mu\,k^\nu
\partial_\alpha g_{\mu \nu}\ .
\label{geophot2}
\end{equation}
Since the considered metric is stationary, $k_0 \equiv - I(\epsilon)k^0$ is a constant, 
according to (\ref{geophot2}).

A free falling observer of coordinates $x^\alpha$ and four-velocity $u^\alpha$ 
can be described by the geodesic equation
\begin{equation}
\frac{du^\alpha}{d\tau}+\Gamma^\alpha_{\mu \nu}\,u^\mu\,u^\nu \ = \ 0.
\label{obsgeod}
\end{equation}
\subsection{Energy of a photon}
Consider a photon of four-wave vector $k^\alpha$ propagating in the background 
metric (\ref{metric}). The general expression of its energy measured by an observer 
of four-velocity $u^\alpha$ reads:
\begin{equation}
E=-g_{\alpha \beta}k^\alpha u^\beta=-k_0u^0-J(\epsilon){\bf k.u}.
\label{energ}
\end{equation}
An expansion of $E$ up to order 3/2 requires a knowledge of $u^0/c$ and $u^i/c$ 
at the same level. On the other hand, $u^i/c$ being a term of order 1/2 
(i.e. $\stackrel{(0)}{u}^i=0$), we need $k^i$ and $J$ up to order 1 only, to 
evaluate $E$. Since the metric components are integer powers of $\epsilon$, 
the functions $I(\epsilon)$ and $J(\epsilon)$ have to be known up to order 1 only,
 in all the developments we will have to consider.

Let us compute $u^0=c.dt/d\tau$. We have
$$c^2d\tau^2=I(\epsilon)c^2dt^2-J(\epsilon)dx^idx^i=\left(I-J{{\bf v}^2\over c^2}\right)c^2dt^2.$$
Let us consider metrics of the form
$$I(\epsilon)=1-2\epsilon+{\cal O}(\epsilon^2)$$
$$J(\epsilon)=1+2\gamma\epsilon+{\cal O}(\epsilon^2)$$
where we recognize the isotropic form of the Schwarzschild metric in the case where the 
Post-Newtonian parameter $\gamma$ is unity. We thus have at the same order
$$d\tau^2=\left(1-2\epsilon-{{\bf v}^2\over c^2}\right)\,dt^2$$
and consequently
$${u^0\over c}=1+\epsilon + {{\bf v}^2\over 2c^2} + {\cal O}(\epsilon^2)$$
We notice that proper and coordinate times differ by terms in $\epsilon$ (Einstein shift) 
and in $v^2/c^2$ (Lorentz time dilation). Now we compute $u^i$
$$
u^i \ = \ \frac{d x^i}{d\tau} \ = \ \frac{d t}{d\tau}\ \frac{d x^i}{dt} = 
  {u^0\over c} v^i
=  \left(1+\epsilon +{{\bf v}^2\over 2c^2}\right)v^i+ {\cal O}(\epsilon^{5/2})
$$
from what we get the energy
\begin{equation}
{E\over c}=-k_0 \left( 1 + \epsilon + {{\bf v}^2\over 2c^2}\right)
-\left[1+(1+2\gamma)\epsilon + {{\bf v}^2\over 2c^2}\right]{{\bf k.v}\over c}.
\label{genenerg}
\end{equation}
\subsection{Light propagation in curved space-time}
The metric tensor is of the form
\begin{equation}
g_{\alpha \beta} \ = \ \eta_{\alpha \beta} +
2 q_\alpha\, \epsilon\, \delta_{\alpha \beta}
\end{equation}
where $q_0 =1$ and $q_i = \gamma$. We have as well
\begin{equation}
g^{\alpha \beta} \ = \ \eta_{\alpha \beta} -
2 q_\alpha\, \epsilon\, \delta^{\alpha \beta}
\end{equation}
The motion of the photon is described by the geodesic equation 
(\ref{geophot1}) or (\ref{geophot2}) and by the null condition (\ref{isotrop}).
\subsubsection{Order 0}

At this order, space-time is flat and the spacecrafts are at rest (no velocities, 
which are 1/2 order terms, and no accelerations, which are first order terms), and 
light propagates according to special relativity. The results of order 0 are obvious, but 
since the same notation will be used in higher order terms, we present here the detailed solution.

Solving eq.(\ref{geophot1}), we find, the $\Gamma$ being first order terms 
$$\stackrel{(0)}{k}^\alpha\ = \ C^\alpha$$
where $C^\alpha$ are four integration constants. From the isotropic condition (\ref{isotrop}),
$$C_0C_0=C^0C^0=C_iC_i=C^iC^i$$
where $C_\alpha\equiv\eta_{\alpha\beta}C^\beta$, so that there is a spatial 
unit vector $n^i\equiv C^i/C^0$ along the direction of the (straight) light ray. 
We can parametrize the order 0 photon worldline using the trivial differential equation
$$\frac{d\stackrel{(0)}{x}^\alpha}{d\lambda}\ =\stackrel{(0)}{k}^\alpha$$
which gives (assuming $\lambda =0$ for the emission time)
$$\stackrel{(0)}{x}^0 \ = \ C^0\, \lambda $$
and if $x^i_{\rm A}$ represents the coordinates of the emitter at emission 
time, we have
$$\stackrel{(0)}{x}^i=x^i_{\rm A}+C^i\lambda$$
Let us remark that interpreting $\stackrel{(0)}{x}^0$ as the Minkowski
 time coordinate $ct$, this could be simply written as
$$\stackrel{(0)}{\bf x}\!\!(t)={\bf x}_{\rm A}+ct.{\bf n}.$$
The spatial coordinate distance $\stackrel{(0)}{r}\!\!(\lambda)$, which 
corresponds to the ``physical instantaneous distance" at this order, is:
\begin{equation}
\stackrel{(0)}{r}\!\!(\lambda )^2=\stackrel{(0)}{x}^i\stackrel{(0)}{x}^i=
\left[C^0 \lambda +{\bf n.x}_{\rm A}\right]^2+K^2
\label{dist0lbd}
\end{equation}
with $K^2=r_{\rm A}^2-({\bf n.x}_{\rm A})^2$ and $r_{\rm A}^2\equiv{\bf x}_{\rm A}^2.$
\subsubsection{Order 1}

$k_0$ being a constant, we obtain $k^0$ at order 1 by
\begin{equation}
\stackrel{(\rightarrow1)}{k}^0=g^{00}\stackrel{(\rightarrow1)}{k}_0=
g^{00}\stackrel{(0)}{k}_0=g^{00}C_0=(1+2\stackrel{(0)}{\epsilon})C^0
\label{k0firstorder}
\end{equation}
where $\stackrel{(0)}{\epsilon}=2r_S/\stackrel{(0)}{r}$. Integrating the spatial 
components of the geodesic equation (\ref{geophot1}), or (\ref{geophot2}), one finds
\begin{equation}
\stackrel{(\rightarrow1)}{k}^i=C^0\left[n^i-(1+\gamma)\stackrel{(0)}{\epsilon}P^i
{\bf n}.\stackrel{(0)}{\bf x}-(\gamma-1)\stackrel{(0)}{\epsilon}n^i\right]
\label{kifirstorder}
\end{equation}
where $P^i$ is defined as
$$P^i\equiv{x^i_{\rm A}-n^i{\bf n.x}_{\rm A}\over K^2}$$
and satisfies ${\bf P.n}=0$. 
Integrating equation
$${d\stackrel{(\rightarrow1)}{x}^\alpha\over d\lambda}=
\stackrel{(\rightarrow1)}{k}^\alpha$$
one finds, in parametric form 
$$\stackrel{(\rightarrow1)}{x}^0\!\!(\lambda)=C^0\lambda+2{GM\over c^2} 
\ln{{\bf n}.\stackrel{(0)}{\bf x}+
\stackrel{(0)}{r}\over{\bf n}.{\bf x}_{\rm A}+r_{\rm A}}$$
$$\stackrel{(\rightarrow1)}{x}^i\!\!(\lambda)=x^i_{\rm A}+C^i\lambda-
(\gamma+1){GM\over c^2}P^i[\stackrel{(0)}{r}-r_{\rm A}]$$
$$-(\gamma-1){GM\over c^2}n^i\ln{{\bf n}.\stackrel{(0)}{\bf x}+
\stackrel{(0)}{r}\over{\bf n.x}_{\rm A}+r_{\rm A}}.$$
Eliminating the affine parameter, one obtains the photon's trajectory in the explicit form
\begin{equation}
\stackrel{(\rightarrow1)}{x}^i(t,n^j)=x^i_{\rm A}+n^ict-(\gamma+1){GM\over c^2}\chi^i(t,n^j)
\label{photontraj}
\end{equation}
with
$$\chi^i(t,n^j)\equiv P^i[\stackrel{(0)}{r}-r_{\rm A}]
+n^i\ln{{\bf n}.\stackrel{(0)}{\bf x}+\stackrel{(0)}{r}\over{\bf n.x}_{\rm A}+r_{\rm A}}$$
where the dependence in the integrations constants $n^i$ is explicitely written 
and  $\stackrel{(0)}{r}$ is given by (\ref{dist0lbd}) with $C^0\lambda$ replaced by $ct$.
\subsubsection{Energy of the photon}

Inserting (\ref{kifirstorder}) in (\ref{genenerg}), one finds
$${1\over C^0}{E\over c}=1-{{\bf n.v}\over c}+{GM\over rc^2}+{{\bf v}^2\over2c^2}$$
$$-\left[(2+\gamma){GM\over\stackrel{(0)}{r}c^2}+{{\bf v}^2\over2c^2}\right]
{{\bf n.v}\over c}$$
\begin{equation}
+(\gamma+1){GM\over c^2}{{\bf P.v}\over c}{{\bf n}.\stackrel{(0)}{\bf x}
\over\stackrel{(0)}{r}}.
\label{enerdevel}
\end{equation}
The first line of the right-hand side (r.h.s.) of (\ref{enerdevel}) contains terms of order 0 
(i.e. $1$), 1/2 (i.e. $-{\bf n.v}/c$) and 1, while terms of order 3/2 are in 
the second and third lines.

Equation (\ref{enerdevel}) shows that knowledge of the energy up to 
order 3/2 requires

- knowledge of $n^i$ up to order 1;

- knowledge of $v^i/c$ up to order 3/2;

- knowledge of $x^i$ up to order 1/2.
\subsubsection{Evaluating the flight time and the constants $n^i$}

At $t=0$, a photon is emitted from $x^i_{\rm A}$ by spacecraft A with
initial position $x^i_{\rm A}$ and velocity $v^i_{\rm A}$, and is received 
at $x^i_{\rm B}(t)$ by spacecraft B with initial position $x^i_{\rm B}$ and 
velocity $v^i_{\rm B}$. We want to compute both the flight time $t$, and 
the constants $n^i$ characterizing this photon.

The equation to solve writes
\begin{equation}
x^i(t,n^j)=x^i_{\rm B}(t)
\label{encounteq1}
\end{equation}
where $x^i(t,n^j)$ represents the photon's trajectory 
(\ref{photontraj}) and $x^i_{\rm B}(t)$ the receiver spacecraft's 
trajectory. Since $i=1,2,3$, equation (\ref{encounteq1}) 
represents three equations involving the four unknown $t$ and $n^i$.
 The fourth equation is the normalization condition 
\begin{equation}
{\bf n}^2=n^in^i=1.
\label{normadir}
\end{equation}
The Taylor development reads
\begin{equation}
x^i_{\rm B}(t)=x^i_{\rm B}+tv^i_{\rm B}+{t^2\over2}
\left[{dv^i_{\rm B}\over dt}\right]_{(t=0)}+...
\label{btrajdevel}
\end{equation}
Since the operator $d/dt=v^i\partial_i$ increases by 1/2 the order of the 
quantity on which it is applied, this development shows that :

- knowledge of $x^i_{\rm B}(t)$ up to 1/2 order requires knowledge of 
$t$ at order 0;

- knowledge of $v^i_{\rm B}(t)=dx^i_{\rm B}(t)/dt$ up to order 3/2 requires 
 knowledge of $t$ at order 1/2.
 
\noindent
Hence, one has to solve equation (\ref{encounteq1}) up to order 1/2 for $t$ and 
up to order 1 for $n^i$.

Let us write (\ref{encounteq1}) in the explicit form, using (\ref{photontraj}) 
and (\ref{btrajdevel}),
\begin{equation}
x^i_{\rm A}+n^ict-(\gamma+1){GM\over c^2}\chi^i(t,n^j)
=x^i_{\rm B}+tv^i_{\rm B}-{t^2\over2}{GMx^i_{\rm B}\over r^3_{\rm B}}
\label{encounteq2}
\end{equation}
where the initial acceleration, given by the geodesic equation (\ref{obsgeod})  
reduces to its newtonian part at order 1. We notice that, due to properties of 
the metric considered (static and diagonal), this is also the expression of the 
initial acceleration up to order 3/2.

At order 0, equations (\ref{encounteq2}) and (\ref{normadir}) read
$$x^i_{\rm A}+\stackrel{(0)}{n}^ic\stackrel{(0)}{t}=x^i_{\rm B}$$
$$\stackrel{(0)}{\bf n}^2=1$$
which give the two (obvious) order 0 solutions
\begin{equation}
c\stackrel{(0)}{t}=\pm\sqrt{(x^i_{\rm B}-
x^i_{\rm A})(x^i_{\rm B}-x^i_{\rm A})}
\label{0ordertime}
\end{equation}
\begin{equation}
\stackrel{(0)}{n}^i=\pm{x^i_{\rm B}-
x^i_{\rm A}\over\sqrt{(x^j_{\rm B}-x^j_{\rm A})(x^j_{\rm B}-x^j_{\rm A})}}
\label{0orderdir}
\end{equation}
where the ``$+$" sign corresponds to a photon emitted by spacecraft A and 
received by spacecraft B, while the ``$-$" sign corresponds to a photon 
emitted by spacecraft B and received by spacecraft A. This is nothing but 
the trivial euclidean prediction in the case where both emitting and receiving 
spacecrafts are at rest.

At order 1/2, using the 0 order equations, equations (\ref{encounteq2}) and 
(\ref{normadir}) read
$$\stackrel{(0)}{n}^i\stackrel{(1/2)}{t}+\stackrel{(0)}{t}\stackrel{(1/2)}{n}^i=
\stackrel{(0)}{t}{v^i_{\rm B}\over c}$$
$$\stackrel{(0)}{\bf n}.\stackrel{(1/2)}{\bf n}=0.$$
The corresponding terms of order 1/2
\begin{equation}
\stackrel{(1/2)}{t}=\stackrel{(0)}{t}{\stackrel{(0)}{\bf n}.{\bf v}_{\rm B}\over c}
={v^i_{\rm B}\over c^2}\left[x^i_{\rm B}-x^i_{\rm A}\right]
\label{1/2ordertime}
\end{equation}
$$\stackrel{(1/2)}{n}^i={v^i_{\rm B}\over c}-\stackrel{(0)}{n}^i{\stackrel{(0)}{\bf n}.
{\bf v}_{\rm B}\over c}$$
are the classical motion and aberration corrections.

The same iterative method provides the next order terms. 
For the energy problem, one only needs $\stackrel{(1)}{n}^i$, 
but the method also gives $\stackrel{(1)}{t}$, which will be useful 
in the time-delay problem
$$
\stackrel{(1)}{t}={1\over2}\left[{{\bf v}_{\rm B}^2\over c^2}+
\left({\stackrel{(0)}{\bf n}.{\bf v}_{\rm B}\over c}\right)^2\right]\stackrel{(0)}{t}+
$$
\begin{equation}
+\stackrel{(0)}{n}^i\left[(1+\gamma){GM\over c^3}\chi^i(\stackrel{(0)}{t},\stackrel{(0)}{n}^j)
-{GMx^i_{\rm B}\over2r^3_{\rm B}c}\stackrel{(0)}{t}^2\right]
\label{1ordertime}
\end{equation}
$$
\stackrel{(1)}{n}^i=-{1\over2c^2}\stackrel{(0)}{n}^i\left[{\bf v}_{\rm B}^2 -
(\stackrel{(0)}{\bf n}.{\bf v}_{\rm B})^2\right]
$$
$$
-{GM\stackrel{(0)}{t}\over2r^3_{\rm B}c}\left[x^i_{\rm B}-
\stackrel{(0)}{n}^i\stackrel{(0)}{\bf n}.{\bf x}_{\rm B}\right]
$$
$$+(\gamma+1){GM\over c^3\stackrel{(0)}{t}}\left[\chi^i-
\stackrel{(0)}{n}^i\stackrel{(0)}{\bf n}.{\bf\chi}\right]_{\left(\stackrel{(0)}{t},
\stackrel{(0)}{n}^k\right)}.
$$
The ``$\ln$" term in $\stackrel{(1)}{t}$, present in the $\chi^i$ term, 
is responsible for the well-known Shapiro time-delay effect.
\section{Time delays}
The time transfer between two satellites A and B is, at order 0,
$$\stackrel{(\rightarrow1)}{t}=\stackrel{(0)}{t}+\stackrel{(1/2)}{t}+\stackrel{(1)}{t}$$
where the contributions $\stackrel{(k)}{t}$ have been calculated in the previous section. 
The 0 order term is the same for both links A$\rightarrow$B and B$\rightarrow$A. 
The first term to be affected by the permutation ${\rm A}\rightleftharpoons{\rm B}$ is 
the 1/2 order term. It is easy to see that, for this permutation, one has
$$(\stackrel{(0)}{t},\stackrel{(0)}{n}^i)\rightleftharpoons(\stackrel{(0)}{t},
-\stackrel{(0)}{n}^i)$$
leading to
$$\stackrel{(1/2)}{t}({\rm A}\rightarrow{\rm B})-
\stackrel{(1/2)}{t}({\rm B}\rightarrow{\rm A})
=\stackrel{(0)}{t}\stackrel{(0)}{n}^i{v^i_{\rm A}+v^i_{\rm B}\over c}$$
which is responsible for the so-called Sagnac effect.
Let us notice that in the TDI problem, one generally needs the emission time as a function 
of the reception time (or, equivalently, the flight time as a function of the reception time), 
rather than the converse. Accordingly, the solution required in simulations generally 
involves the order 0 solution (\ref{0ordertime}),(\ref{0orderdir}) with the minus sign.

\section{Einstein frequency shifts}
Three identical clocks, one aboard each LISA spacecraft, beat at a common proper frequency. 
The quantity measured on spacecraft B is the (relative) difference between the frequency 
received from spacecraft A, and the proper frequency measured on B. For a photon starting 
from A at $t=0$, and received at B at time $t$, this relative frequency shift writes
\begin{equation}
z_{({\rm B}\leftarrow{\rm A})}(t_{\rm A}=0)={E_{\rm B}(t)\over E_{\rm A}(t=0)}-1
\label{z}
\end{equation}
since the nominal frequency of the oscillator aboard B is identical to the one of the 
oscillator aboard A, which is nothing but the initial energy of the photon emitted from A. 

Let us remark that $1+z$ as defined in (\ref{z}) corresponds to the standard definition
of a frequency shift (frequency at reception over frequency at emission) in the case 
where the solution (\ref{0ordertime},\ref{0orderdir}) with the $+$ sign is considered. In 
the other case ($-$ sign), $1+z$ as defined in (\ref{z}) corresponds to the inverse
ratio (frequency at emission over frequency at reception). 
\subsection{Contribution of the different orders}
The development of the expression for the energy up to order 3/2 is given by (\ref{enerdevel}). 
This expression requires the velocity of the spacecraft up to order 3/2
$$\stackrel{(3/2)}{v}^i_{\rm B}(t)=
v^i_{\rm B}+t{dv^i_{\rm B}\over dt}+{t^2\over2}{d^2v^i_{\rm B}\over dt^2}$$
$$=v^i_{\rm B}-tGM{x^i_{\rm B}\over r_{\rm B}^3}
-{t^2\over2}GM\left[{v^i_{\rm B}\over r_{\rm B}^3}
-3{x^i_{\rm B}{\bf x}_{\rm B}.{\bf v}_{\rm B}\over r^5_{\rm B}}\right].$$
Substituting the above expression in equation (\ref{enerdevel}), one finds, writing 
$\Delta E\equiv E_{\rm B}(t)-E_{\rm A}(t=0)$
\begin{equation}
\Delta E=\stackrel{(1/2)}{\Delta E}+\stackrel{(1)}{\Delta E}+\stackrel{(3/2)}{\Delta E}
\label{deltaEdevel}
\end{equation}
with, writing $N\equiv\Delta E/(cC^0)$,
\begin{equation}
\stackrel{(1/2)}{N}=-\stackrel{(0)}{n}^i{v^i_{\rm B}-v^i_{\rm A}\over c}
\label{X1/2}
\end{equation}
$$\stackrel{(1)}{N}={v^i_{\rm B}-v^i_{\rm A}\over c^2}
\left[\stackrel{(0)}{n}^i\stackrel{(0)}{\bf n}.{\bf v}_{\rm B}
-{1\over2}(v^i_{\rm B}-v^i_{\rm A})\right]$$
\begin{equation}
+{GM\over c^2}\left[{1\over r_{\rm B}}-{1\over r_{\rm A}}\right]
+{GM\over c}\stackrel{(0)}{t}{\stackrel{(0)}{\bf n}.{\bf x}_{\rm B}\over r^3_{\rm B}}
\label{X1}
\end{equation}
$$\stackrel{(3/2)}{N}=-\stackrel{(1)}{n}^i{v^i_{\rm B}-v^i_{\rm A}\over c}
-\stackrel{(0)}{n}^i{v^i_{\rm B}{\bf v}_{\rm B}^2-v^i_{\rm A}{\bf v}_{\rm A}^2\over2c^3}$$
$$-{GM\over c^2}\stackrel{(0)}{t}{{\bf v}_{\rm B}.{\bf x}_{\rm B}\over r^3_{\rm B}}
+{GM\over2c}\stackrel{(0)}{n}^i\stackrel{(0)}{t}^2\left[{v^i_{\rm B}\over r_{\rm B}^3}
-3{x^i_{\rm B}{\bf x}_{\rm B}.{\bf v}_{\rm B}\over r^5_{\rm B}}\right]$$
$$-(2+\gamma){GM\over c^3}\stackrel{(0)}{n}^i\left[{{v^i_{\rm B}}\over r_{\rm B}}
-{{v^i_{\rm A}}\over r_{\rm A}}\right]$$
$$+(1+\gamma){GM\over c^3}\stackrel{(0)}{P}^i\stackrel{(0)}{n}^j
\left[{x^j_{\rm B}v^i_{\rm B}\over r_{\rm B}}-{x^j_{\rm A}v^i_{\rm A}\over r_{\rm A}}\right]$$
where $\stackrel{(0)}{P}^i\equiv P^i\left(n^i=\stackrel{(0)}{n}^i\right)$. 
Since (\ref{deltaEdevel}) has no 0 order term, one only needs $E_{\rm a}$ up to 
order 1 to obtain $z$ up to order 3/2. Setting $D\equiv E_{\rm A}(t=0)/(cC^0)$ yields
$$\stackrel{(0)}{D}=1$$
\begin{equation}
\stackrel{(1/2)}{D}=-{\stackrel{(0)}{\bf n}.{\bf v}_{\rm A}\over c}
\label{Y1/2}
\end{equation}
$$\stackrel{(1)}{D}=-{v^i_{\rm A}\over c^2}(v^i_{\rm B}-
\stackrel{(0)}{n}^i\stackrel{(0)}{\bf n}.{\bf v}_{\rm B})+
{{\bf v}_{\rm A}^2\over2c^2}+{GM\over r_{\rm A}c^2}.$$
Hence, one finds
\begin{equation}
\stackrel{(1/2)}{z}=\stackrel{(1/2)}{N}
\label{z1/2}
\end{equation}
\begin{equation}
\stackrel{(1)}{z}=\stackrel{(1)}{N}-\stackrel{(1/2)}{N}\stackrel{(1/2)}{D}
\label{z1}
\end{equation}
$$\stackrel{(3/2)}{z}=\stackrel{(3/2)}{N}-\stackrel{(1)}{N}\stackrel{(1/2)}{D}-
\stackrel{(1/2)}{N}\stackrel{(1)}{D}+\stackrel{(1/2)}{N}\left[\stackrel{(1/2)}{D}\right]^2.$$
It appears that $\stackrel{(1/2)}{z}$ is unchanged under the permutation ${\rm A}
\rightleftharpoons{\rm B}$, as it should be, since the Doppler effect only depends on 
relative motion in the framework of special relativity. Of course, this is not the case 
as soon as gravity is acting, i.e. in first and higher order terms. Indeed, one finds
$$\stackrel{(1)}{z}_{({\rm B}\leftarrow{\rm A})}(t_{\rm A}=0)-\stackrel{(1)}
{z}_{({\rm A}\leftarrow{\rm B})}(t_{\rm B}=0)$$
$$={2GM\over c^2}\left[{1\over r_{\rm B}}-{1\over r_{\rm A}}\right]
+{GM\over c}\stackrel{(0)}{n}^i\stackrel{(0)}{t}
\left[{x^i_{\rm A}\over r^3_{\rm A}}+{x^i_{\rm B}\over r^3_{\rm B}}\right].$$
The first term originates in the Einstein gravitational Doppler effect, 
while the second is related to the accelerations of spacecrafts in the 
gravitational field of the Sun.
\subsection{Orders of magnitude for the LISA configuration}
One is tempted to derive an order of magnitude for the contributions to the frequency shift. 
Indeed, one has, from (\ref{epsilondef})
$$\epsilon\sim10^{-8}$$
and the difference between the distance of each station to the Sun satisfies
$${\delta r\over r}\leq{L\over r} \cos{\pi\over3}\sim{1\over60}$$
where $L\sim5.10^6$ km is the typical interspacecraft distance. An estimation of the 
terms of orders 1/2, 1 and 3/2 leads to
$$\stackrel{(1/2)}{z}\sim\delta\left(\epsilon^{1/2}\right)\sim\epsilon^{1/2}
{\delta r\over r}\sim2.10^{-6}$$
$$\stackrel{(1)}{z}\sim\delta\left(\epsilon\right)\sim\epsilon{\delta r\over r}
\sim2.10^{-10}$$
$$\stackrel{(3/2)}{z}\sim\delta\left(\epsilon^{3/2}\right)\sim\epsilon^{3/2}
{\delta r\over r}\sim2.10^{-14}.$$

It is interesting to point out that, due to the peculiar LISA configuration, terms 
of orders 1/2 and 1 are actually considerably smaller than the above estimates. 
Let us evaluate these terms more carefully. 

\noindent
In the ideal case, where LISA spacecraft form a perfect equilateral triangle, velocities of spacecrafts A and B have the same projection on 
vector $\stackrel{(0)}{\bf n}$, such that from (\ref{z1/2}),
$$\stackrel{(1/2)}{z}=0.$$
In fact, the above cancellation is obtained as soon as the orbits about the center of 
mass of the satellite constellation are coplanar and circular, 
with a velocity proportional to 
the center of mass distance. Since this property is not exactly verified by the real configuration, 
the value of $\stackrel{(1/2)}{z}$ is reduced by a factor $\sim L/r$, i.e. $\stackrel{(1/2)}{z}\sim7.10^{-8}$. 
At this order, the frequency shift is due to the arm length variations (flexing).

Let ${\bf v}_{\rm AB}\equiv{\bf v}_{\rm B}-{\bf v}_{\rm A}$. 
From (\ref{z1}), (\ref{X1}), (\ref{X1/2}) and (\ref{Y1/2}), the term of order 1 writes
$$\stackrel{(1)}{z}=\left(\stackrel{(0)}{\bf n}.{\bf v}_{\rm AB}\over c\right)^2
-{1\over2}\left({\bf v}_{\rm AB}\over c\right)^2$$
\begin{equation}
+{GM\over c^2}\left({1\over r_{\rm B}}-{1\over r_{\rm A}}\right)
+{GM\over c}\stackrel{(0)}{t}{\stackrel{(0)}{\bf n}.{\bf x}_{\rm B}\over r^3_{\rm B}}.
\label{z1devel}
\end{equation}
The first term on the rhs is small, $\sim5.10^{-15}$, since it is nothing but $\stackrel{(1/2)}{z}^2$. 
\newline In the second term, ${\bf v}_{\rm AB}^2$ turns out to be three times the squared 
modulus of the relative velocity of a spacecraft with respect to LISA's center of 
mass, i.e. $VL/(r\sqrt{3})$ with $V$ the orbital velocity of this center. Hence,
\begin{equation}
{1\over2}\left({\bf v}_{\rm AB}\over c\right)^2\sim{V^2\over c^2}{L^2\over2r^2}\sim6.10^{-12}.
\label{z1rhs2}
\end{equation}
The third term of the rhs of (\ref{z1devel}) is of order
$${GM\over rc^2}{\delta r\over r}\sim2.10^{-10}$$
while the fourth writes
$${GM\over c^2}c\stackrel{(0)}{t}{\stackrel{(0)}{\bf n}.{\bf x}_{\rm B}\over r^3_{\rm B}}
\sim{GM\over rc^2}{L\over r}\cos\theta\sim2.10^{-10}$$
where $\theta$ spans between $\pi/3$ and $2\pi/3$. 
\newline The point is that the third and fourth rhs terms of equation (\ref{z1devel}), of order $(GM/c^2)(L/r^2)$ each, nearly cancel. Indeed, the euclidean triangular assumption,
$$r_{\rm A}^2=r_{\rm B}^2-2r_{\rm B}{\bf L}.\widehat{\bf x}_{\rm B}+L^2$$
with ${\bf x}_{\rm A,B}=r_{\rm A,B}\ \widehat{\bf x}_{\rm A,B}$ and 
${\bf L}=c\stackrel{(0)}{t}\stackrel{(0)}{\bf n}$ leads to
$${1\over r_{\rm A}}={1\over r_{\rm B}}+{{\bf L}.\widehat{\bf x}_{\rm B}\over r_{\rm B}^2}
+\left[-1+3\left(\stackrel{(0)}{\bf n}.\widehat{\bf x}_{\rm B}\right)^2\right]
{L^2\over 2r_{\rm B}^3}+{\cal O}\left({L^3\over r^4}\right).$$
Since $\left(\stackrel{(0)}{\bf n}.\widehat{\bf x}_{\rm B}\right)^2\leq1/4$, 
the sum of the third and fourth terms in (\ref{z1devel}) is about
\begin{equation}
{1\over2}{GM\over rc^2}\left({L\over r}\right)^2\sim6.10^{-12}.
\label{term3+term4}
\end{equation}
Furthermore, it turns out that the residual (third+fourth)-term in (\ref{z1devel}), of order $(GM/c^2)(L^2/r^3)$, nearly cancels with the second term in (\ref{z1devel}). Indeed, at first order, each spacecraft obeys Newton's law
$${d{\bf v}\over dt}=-GM{{\bf x}\over r^3}$$
so that their relative motion is described by
$${d{\bf v}_{\rm AB}\over dt}=-GM\left({{\bf L}\over r^3}-
3{({\bf r.L}){\bf r}\over r^5}\right)+
{\cal O}\left({L^2\over r^4}\right)$$
with ${\bf v}_{\rm AB}=d{\bf L}/dt$. 
Since $d\left({\bf L}.{\bf v}_{\rm AB}\right)={\bf L}.d{\bf v}_{\rm AB}+
{\bf v}_{\rm AB}^2dt$, it turns out that the sum of the second, third and fourth 
terms of (\ref{z1devel}) reduces to
\begin{equation}
-{1\over2c^2}{d\over dt}\left(L\stackrel{(0)}{\bf n}.{\bf v}_{\rm AB}\right)+
{\cal O}\left({L^3\over r^4}\right)
\label{z1rhs3+4}.
\end{equation}
The order of magnitude of the ${\cal O}(L^3/r^4)$ terms neglected at the different steps leading to (\ref{z1rhs3+4}) is
$${GM\over c^2}{L^3\over r^4}\cos\theta\sim2.10^{-13}.$$
But the first term of (\ref{z1rhs3+4}) is even smaller, since
$${1\over2c^2}{d\over dt}\left(L\stackrel{(0)}{\bf n}.{\bf v}_{\rm AB}\right)\sim
{1\over2c}L{\stackrel{(1/2)}{z}\over1\ {\rm yr}}\sim3.10^{-14}.$$
Consequently, $\stackrel{(1)}{z}$ appears to be of order $2.10^{-13}$, i.e. only an order of magnitude larger than $\stackrel{(3/2)}{z}$.

In the next section, we will confirm the partial cancelling of $\stackrel{(1/2)}{z}$ 
and $\stackrel{(1)}{z}$ found analytically, thanks to a numerical relativistic 
simulation of LISA's optical links.

\section{Numerical results}
\begin{figure*}
\centering
\includegraphics[width=1.3\columnwidth]{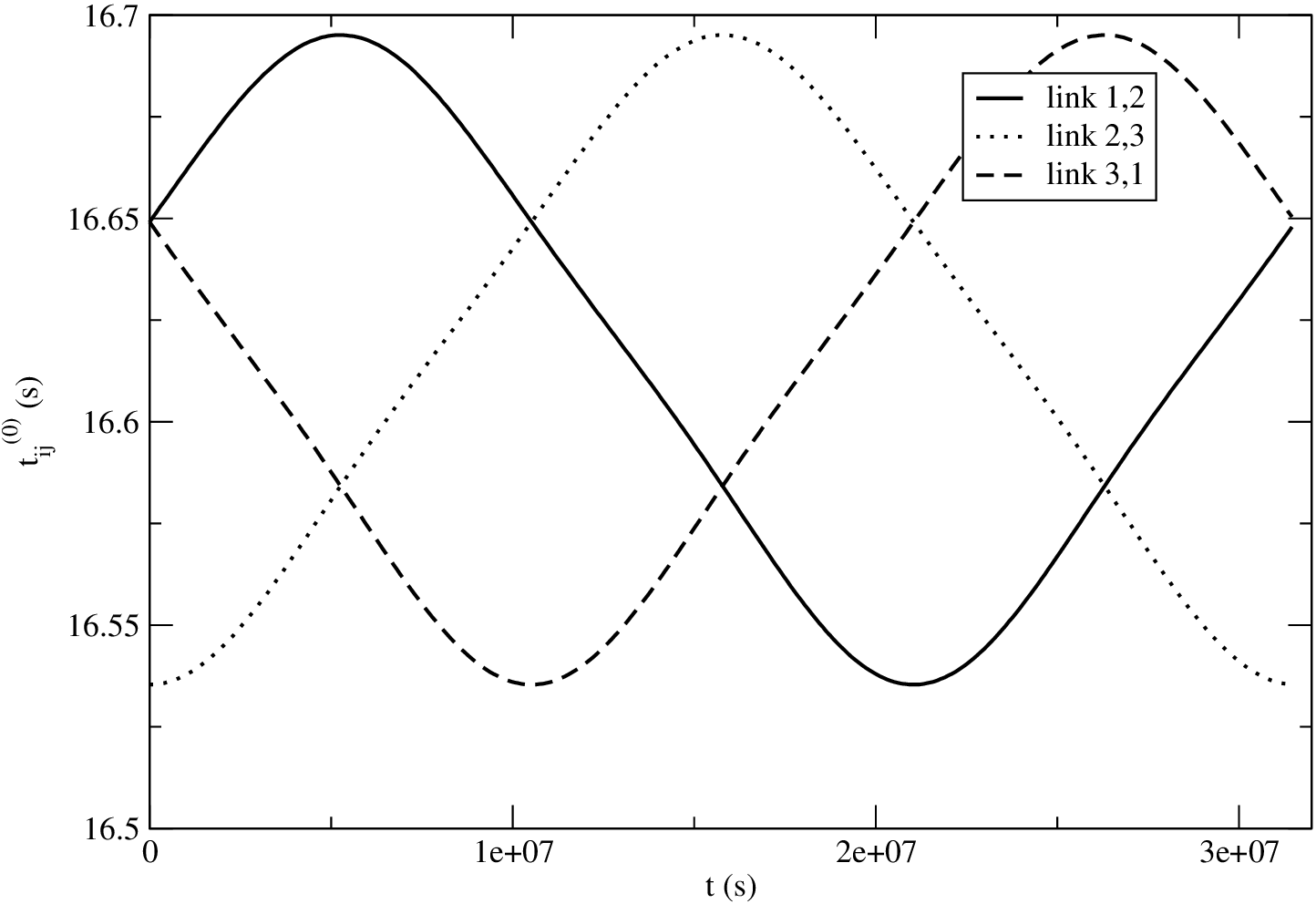}
\caption{\label{fig1} Annual variations of the time delays (zeroth order) along the
LISA arms.
Solid line: $\stackrel{(0)}{t_{12}}$, short dashed line: $\stackrel{(0)}{t_{23}}$, 
long dashed line: $\stackrel{(0)}{t_{31}}$.}
\end{figure*}

\begin{figure*}
\centering
\includegraphics[width=1.3\columnwidth]{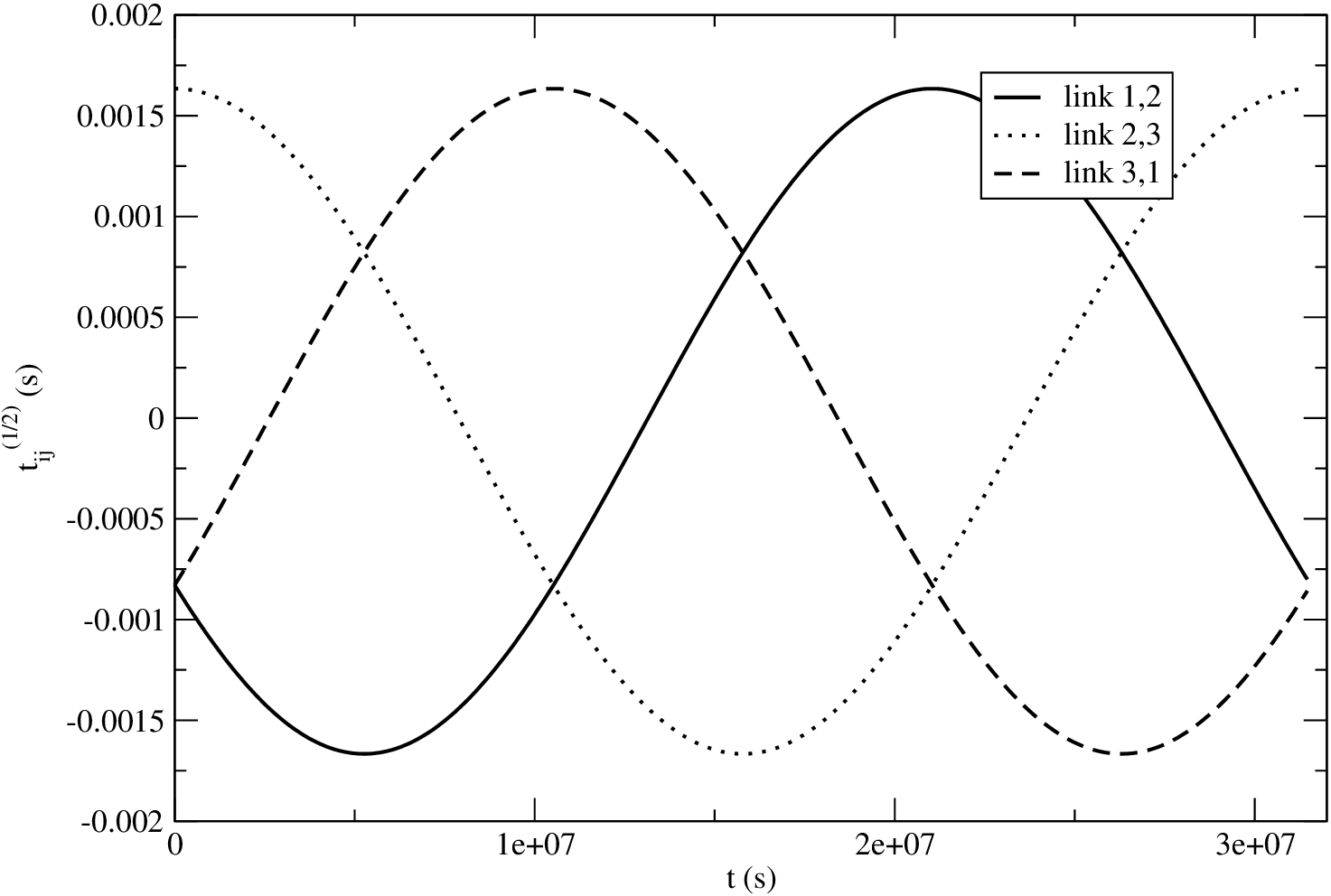}
\caption{\label{fig2} Annual variations of the contribution of order 1/2
to the time delays. Solid line:
$\stackrel{(1/2)}{t_{12}}$, short dashed line: $\stackrel{(1/2)}{t_{23}}$, long dashed line: 
$\stackrel{(1/2)}{t_{31}}$.
}
\end{figure*}

\begin{figure*}
\centering
\includegraphics[width=1.3\columnwidth]{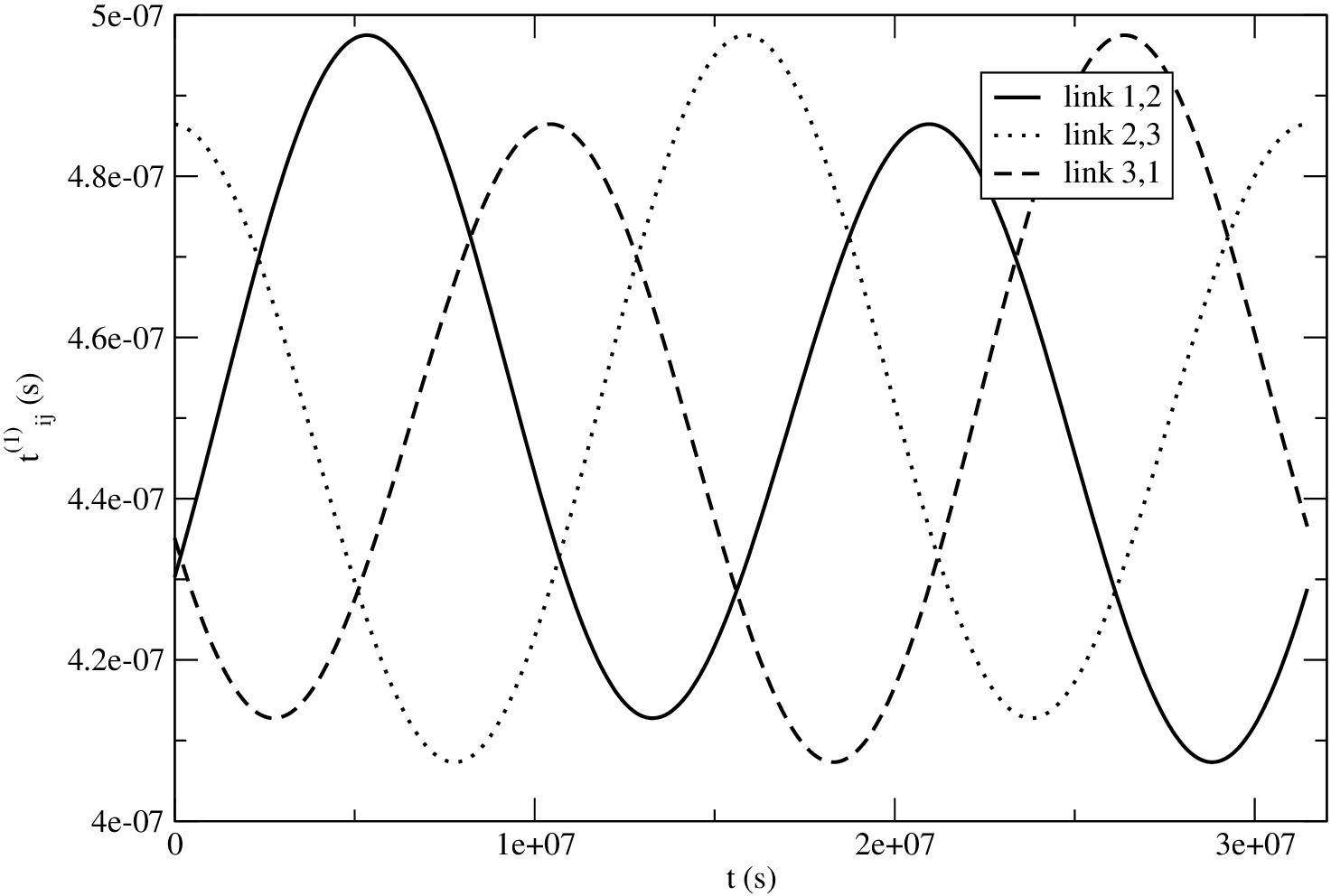}
\caption{\label{fig3} Annual variations of the contribution of order 1
to the time delays. Solid line:
$\stackrel{(1)}{t_{12}}$, short dashed line: $\stackrel{(1)}{t_{23}}$, 
long dashed line: $\stackrel{(1)}{t_{31}}$. }
\end{figure*}

\begin{figure*}
\centering
\includegraphics[width=1.3\columnwidth]{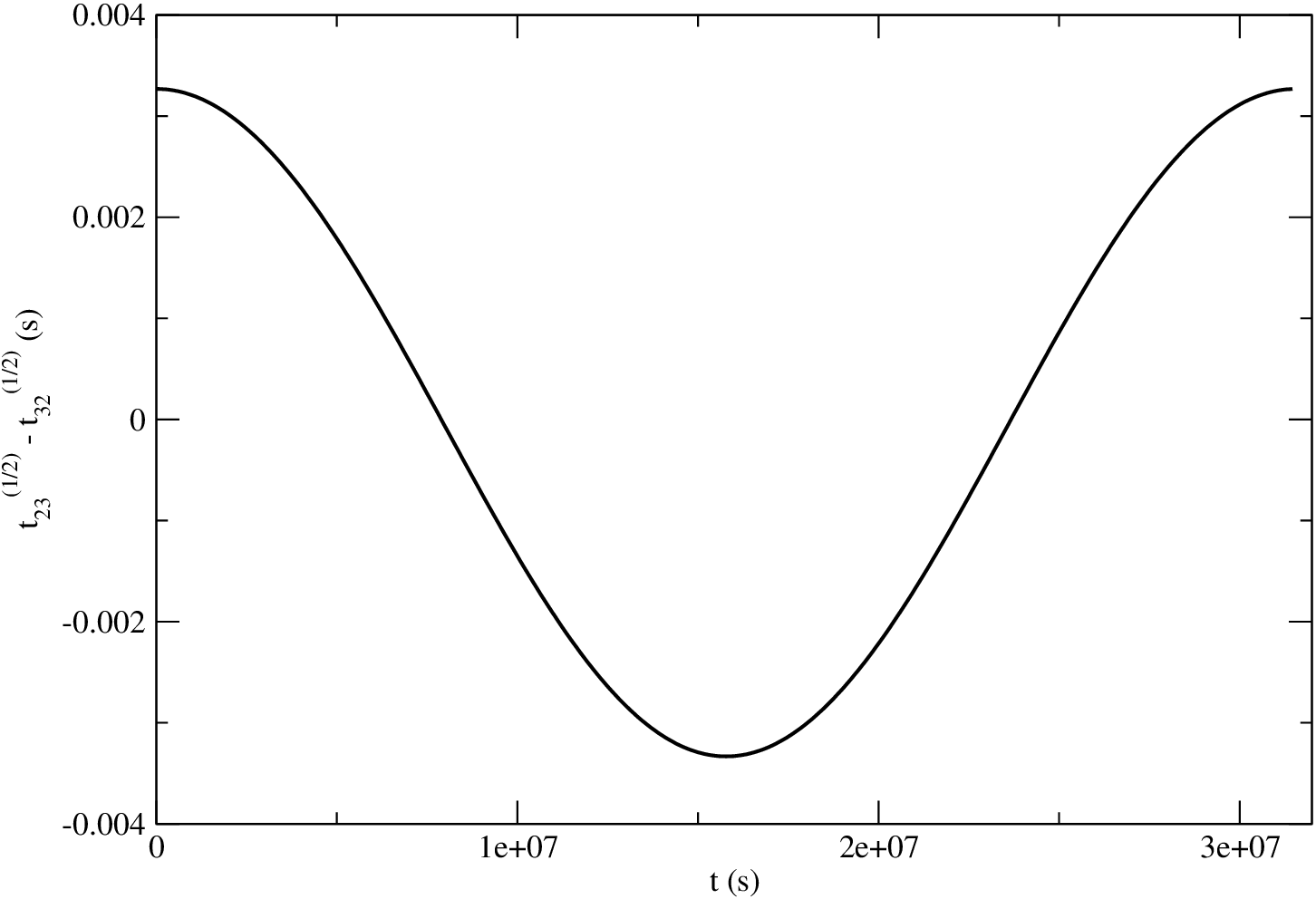}
\caption{\label{fig4} Annual variations of the Sagnac differential time
between spacecraft 2 and spacecraft 3.}
\end{figure*}

\begin{figure*}
\centering
\includegraphics[width=1.3\columnwidth]{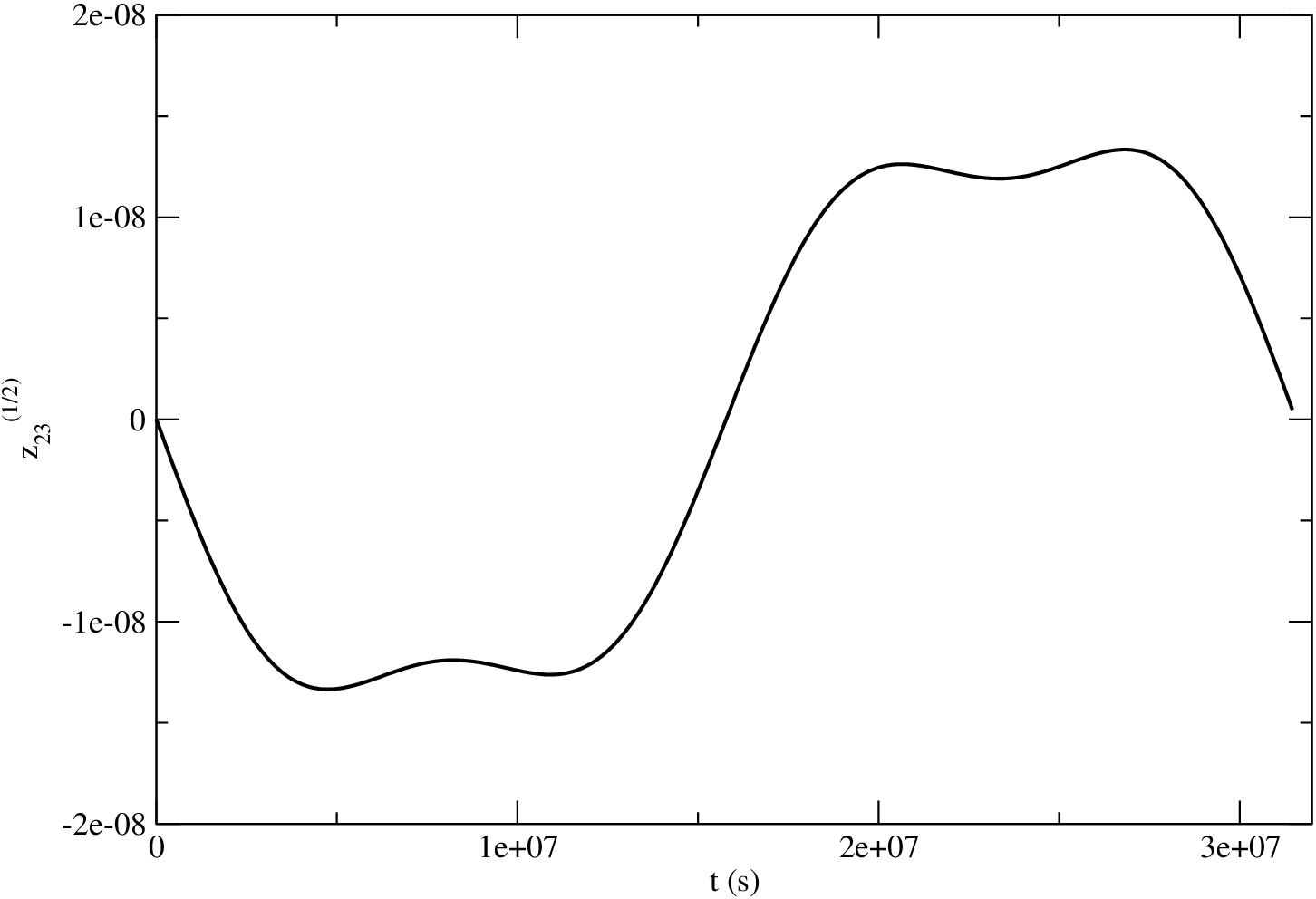}
\caption{\label{fig5} Annual variations of the contribution of order 1/2
to the frequency shift.}
\end{figure*}

\begin{figure*}
\centering
\includegraphics[width=1.3\columnwidth]{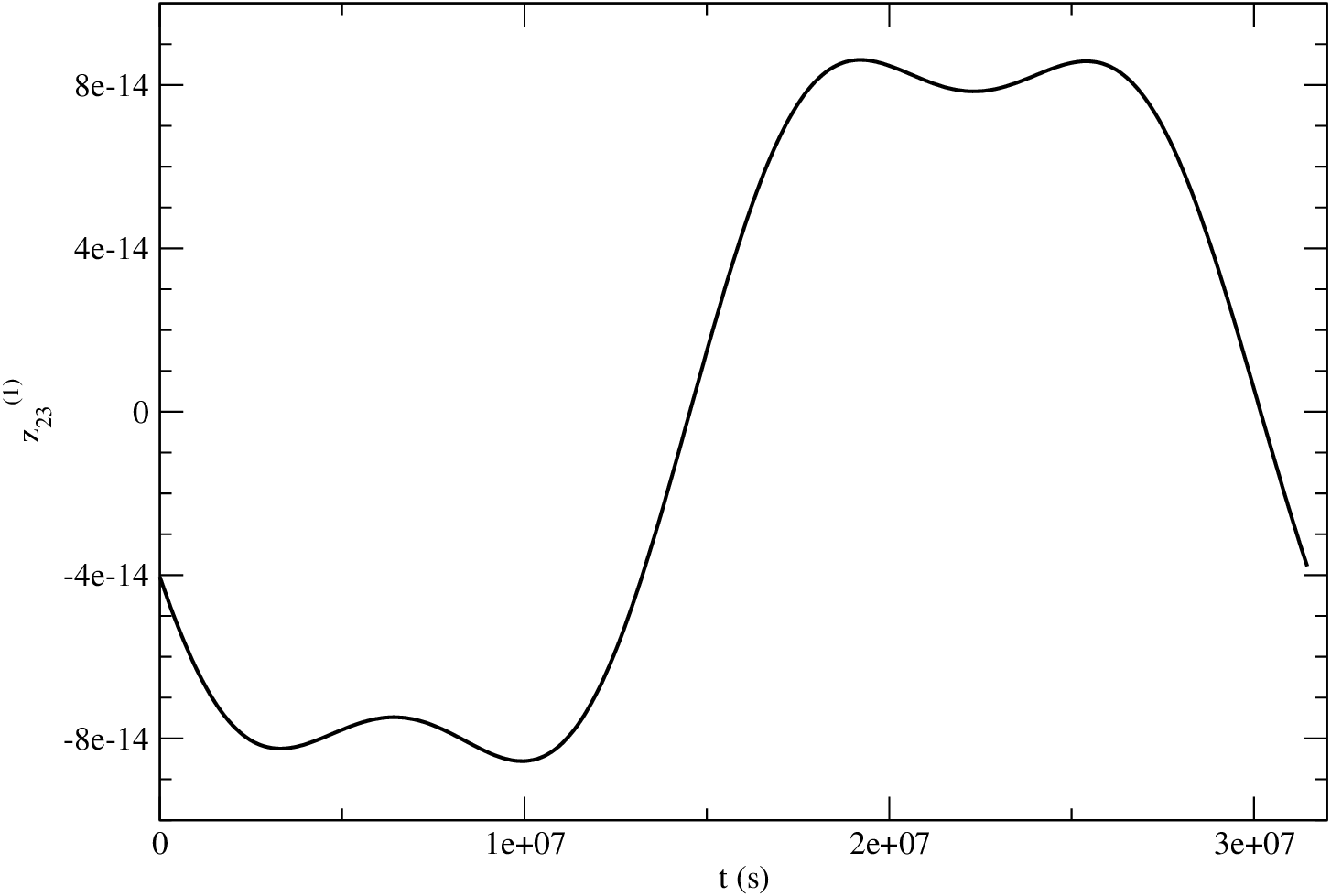}
\caption{\label{fig6} Annual variations of the contribution of order 1
to the frequency shift.}
\end{figure*}

In order to evaluate the orders of magnitude of the various terms of the
expansions in the time delays and in the frequency shifts, we have made
a toy model in which the spacecraft orbits are classical (Keplerian).
At any date, the link between spacecraft A and B is assumed starting from
the Keplerian location of A at current time $t_A$ and joining B at a place
deduced from the keplerian position of B at time $t_A+t$, through a relativistic
continuation. It is known (this has been discussed for instance recently in \cite{DNV}),
that the three spacecraft can be kept at approximately constant mutual distances,
with the constellation's center of mass on a circular orbit of 1 A.U. radius,
at the condition that each orbit has a small eccentricity $e$, a small orbital
inclination $\mu $ on the ecliptic, with the semimajor axes of spacecraft 2 and 
spacecraft 3 orbits shifted by 120 and 240 degrees respectively from spacecraft 1's. 
In fact, if the three spacecraft are required to be equivalent
(i.e., same orbit eccentricity $e$ and inclination angle $\mu$), at
first order in $e$, the distances between
spacecraft are constant and the angle of the LISA plane with respect to
the ecliptic, $\nu$, is 60 degrees. However, exact numerical
investigations show that second order terms in $e$ give rise to
important variations of the interferometric arm lenghts ("breathing")
and to a large Doppler effect. The choice of the orbital parameters
($e$, $\nu$)influences the "breathing" amplitude. 
A carefull choice of the orbital parameters
can reduce the "breathing" amplitude from about 120,000 km (for a naive
choice of $\nu$=60 degrees) to less than 50,000 km. 
Such orbital parameters were chosen in existing simulation
codes \cite{Cornish},\cite{Vallisneri}, and it has been shown in
reference \cite{NKVD} that this choice is the optimal one.
We summarize herafter the results of \cite{NKVD} in terms of
parametrized orbits.
The angle $\nu$ is defined as
$$
\nu \ = \ \frac{\pi }{3}+ \frac{5}{8}\, \alpha 
$$
where $\alpha\equiv L/2R$. $L$ is
the nominal interspacecraft distance ($5\,10^6$ km) and $R$ the radius of the
circular LISA orbit (1 a.u.), so that $\alpha \sim 1/60$ is a small parameter.
This small change of inclination is responsible for a large reduction
of flexing with respect to the strict 60 degrees often quoted in the 
literature \cite{DNV}. 
The inclination $\mu$ of the
orbital planes with respect to the ecliptic is:
$$
\tan\mu \ = \  \frac{\alpha \sin\nu}{\sin(\pi /3)+\alpha \cos\nu }
$$
 Though the center of mass keeps on a
circular orbit, the spacecraft stay on slightly elliptical orbits
(numbered $b=1,2,3$). Their
common eccentricity is:
$$
e \ = \ \left[1+\frac{4\alpha \cos\nu }{\sqrt{3}}+ \frac{4 \alpha^2}{3} \right]^{1/2}
-1
$$
In order to parametrize the orbital motions by the time, we need
the eccentric anomalies $\Psi_b$, that are functions of the time $t$,
defined via the implicit
equations:
$$
\Psi_b - e\sin\Psi_b \ = \ \Omega t - \theta_b
$$
involving a phase shift $\theta_b \equiv 2(b-1)\pi/3$.
Owing to the smallness of $e$, the numerical solution is very fast.
Then, let us set
$$
\left\{
\begin{array}{l}
X_b \ = \ R\,(\cos\Psi_b - e)\,\cos\mu \\
Y_b \ = \ R\,\sqrt{1-e^2}\,\sin\Psi_b \\
Z_b \ = \ - R\,(\cos\Psi_b - e)\,\sin\mu
\end{array}
\right.
$$
Finally, the positions of the three spacecraft at time $t$ are
deduced from the preceding ones via rotations in the $(X,Y)$-plane:
\begin{equation}
\left\{
\begin{array}{l}
x_b \ = \ X_b\ \cos \theta_b - Y_b\ \sin \theta_b \\
y_b \ = \ X_b \ \sin \theta_b + Y_b\ \cos \theta_b  \\
z_b \ = \ Z_b
\end{array}
\right.
\label{orbits}
\end{equation}
At this point, we have three Keplerian orbits, with parameters such that
the distances between test masses are constant at first order in $\alpha$,
but actually varying by about 1\% (about 50,000 km) peak-to-peak during the year. The
spacecraft velocities are computed according to:
\begin{equation}
\left\{
\begin{array}{l}
\dot{x}_b \ = \ \dot{X}_b\ \cos \theta_b - \dot{Y}_b\ \sin \theta_b \\
\dot{y}_b \ = \ \dot{X}_b\ \sin \theta_b + \dot{Y}_b\ \cos \theta_b  \\
\dot{z}_b \ = \ \dot{Z}_b
\end{array}
\right.
\label{veloc}
\end{equation}
with:
$$
\left(
\begin{array}{c}
\dot{X}_b \\
\dot{Y}_b \\
\dot{Z}_b 
\end{array}
\right) \ = \ \frac{\Omega R}{1-e\cos\Psi_b} \ 
\left(
\begin{array}{c}
 -\cos\mu\, \sin\Psi_b   \\
 \sqrt{1-e^2}\,\cos\Psi_b \\
 \,\sin\mu\,\sin\Psi_b
\end{array}
\right)
$$
Once the positions and velocities are known, it is easy to implement the 
time delays up to order 1. Fig.\ref{fig1},\ref{fig2},\ref{fig3} show the
contributions of orders 0, 1/2 and 1 respectively.
We can also evaluate the Sagnac term as the difference between the 
propagation time from spacecraft 2 to 3 and the time from 3 to 2:
see Fig.\ref{fig4}. Remark that the maximum amplitude of this 
differential time delay is approximately twice the maximum
amplitude of the $\stackrel{(1/2)}{t}$ contribution, because
there is a change of sign. Annual variations of the frequency 
shift at order 1/2 are represented on
Fig.\ref{fig5}. The contribution of order 1 is shown on Fig.\ref{fig6}.
\section{Conclusion}
We have given the formulas that allow one to compute the time delays to be
used
in a numerical model of LISA knowing the Keplerian orbits of the
spacecraft. At the lowest order, we have the well-known ``flexing''of the 
triangle, of (peak to peak) amplitude slightly less than 48,000 km. At the next order, 
we find an extra 
correction of
amplitude about 960 km, larger than the allowed error for phase noise cancellation
after TDI. The last computed order is negligible (less than 30 m). The
interest of the derivation is that it includes naturally all relativistic
effects (for instance the Sagnac effect) introduced up to now by ``ad hoc'' 
considerations.

We have also estimated the global frequency shift due to the various motions 
and to propagation in the gravitational field of the Sun. We have
found the rather surprising result that the General Relativistic contribution is much
smaller than a priori expectations, only due to the LISA configuration. 
The global frequency shift, reduced to the residual relative motions of the
spacecraft, is of the order of 8 MHz (peak to peak).

\section{Acknowledgement}
S. Pireaux acknowledges a CNES postdoctoral grant and a convention CNES-INSU 02/CNES/282.

\end{document}